\documentclass[10pt,aps,prb,superscriptaddress,nobibnotes,notitlepage,twocolumn]{revtex4-2}

\usepackage{mathptmx}       
\usepackage{mathtools}      
\usepackage{bm}            

\usepackage{graphicx}       

\usepackage{color}          

\usepackage[normalem]{ulem}

\usepackage{multirow}

\usepackage{tabularx}

\usepackage{hyperref}       
\hypersetup{colorlinks=true, linkcolor=blue, citecolor=blue, urlcolor=blue}

\newcommand{\beq}[1]{\begin{equation}\label{#1}} 
\newcommand{\eeq}{\end{equation}}                
\newcommand{\eep}{\;.\end{equation}}             
\newcommand{\eec}{\;,\end{equation}}             

\newcommand{\lb}{\left(}
\newcommand{\rb}{\right)}

\newcommand{\g}{\gamma}

\newcommand{\la}{\lambda}
\renewcommand{\th}{\theta}

\newcommand{\D}{\Delta}
\newcommand{\G}{\Gamma}

\DeclareMathAlphabet{\mathcal}{OMS}{cmsy}{m}{n} 

\newcommand{\C}{\mathcal{C}}   
\newcommand{\M}{\mathcal{M}}   
\newcommand{\N}{\mathcal{N}}   

\newcommand{\V}{\mathcal{V}}

\newcommand{\bvec}[1]{\mathbf{#1}}

\newcommand{\x}{\bvec{x}}
\newcommand{\kv}{\bvec{k}}
\renewcommand{\u}{\bvec{u}}
\newcommand{\gv}{\bvec{g}}

\newcommand{\av}{\bvec{a}}

\newcommand{\rv}{\bvec{r}}

\newcommand{\VGSFE}{\mathcal{V}_{\mathrm{GSFE}}}
\newcommand{\Velastic}{\mathcal{V}_{\mathrm{elastic}}}

\newcommand*\dd{\mathop{}\!\mathrm{d}} 

\newcommand{\kp}{\bvec{k}\cdot\bvec{p}}

\newcommand{\Gfp}{\Gamma_{0+}}    
\newcommand{\Gfm}{\Gamma_{0-}}    
\newcommand{\Gfpm}{\Gamma_{0\pm}} 
\newcommand{\Gdm}{\Gamma_{1-}}    
\newcommand{\Gddm}{\Gamma_{2-}}   

\newcommand{\Kdp}{\mathrm{K}_{1+}}               
\newcommand{\Kddp}{\mathrm{K}_{2+}}              

\newcommand{\HarvardQSE}{Quantum Science and Engineering, Harvard University, Cambridge, Massachusetts 02138, USA}
\newcommand{\HarvardPhysics}{Department of Physics, Harvard University, Cambridge, Massachusetts 02138, USA}
\newcommand{\HarvardSeas}{John A.~Paulson School of Engineering and Applied Sciences, Harvard University, Cambridge, Massachusetts 02138, USA}
\newcommand{\FSU}{Department of Physics, Florida State University, Tallahassee, Florida 32306, USA}
\newcommand{\NTU}{School of Electrical and Electronic Engineering, Nanyang Technological University Singapore, 50 Nanyang Avenue, 639798, Singapore}

\begin{document}

\title{
Twisted bilayer graphene from first-principles:
structural and electronic properties
}

\author{Albert Zhu}
\affiliation{\HarvardQSE}

\author{Daniel Bennett}
\email{daniel.bennett@ntu.edu.sg}
\affiliation{\HarvardSeas}
\affiliation{\NTU}

\author{Daniel T.~Larson}
\affiliation{\HarvardPhysics}

\author{Mohammed M. Al Ezzi}
\affiliation{\HarvardSeas}

\author{Efstratios Manousakis}
\affiliation{\FSU}

\author{Efthimios Kaxiras} 
\affiliation{\HarvardSeas}
\affiliation{\HarvardPhysics}

\begin{abstract}
We present a comprehensive first-principles study of twisted bilayer graphene (tBLG) for a wide range of twist angles, with a focus on structural and electronic properties.
By employing density functional theory (DFT) with an optimized local basis set, we simulate tBLG, obtaining fully relaxed commensurate structures for twist angles down to 0.987°.
For all angles the lattice relaxation agrees well with continuum elastic models.
For angles accessible to plane-wave DFT ({\sc vasp}), we provide a detailed comparison with our local basis DFT ({\sc siesta}) calculations, demonstrating excellent agreement in both the atomic and electronic structure.
The dependence of the Fermi velocity and band width on the twist angle shows qualitative agreement with results from an `exact' $\kp$ continuum model, but reveals a small twist angle offset.
Additionally, we provide details of the low-energy wavefunction character, band inversion and symmetries.
Our results provide an \textit{ab initio} reference point for the microscopic structure and electronic properties of tBLG which will serve as the foundation for future studies incorporating many-body effects.
\end{abstract}

\maketitle


\section{Introduction
\label{sec:intro}
}

Twisted bilayer graphene (tBLG) has been shown to host a variety of exotic phenomena, such as superconductivity \cite{cao2018unconventional,lu2019superconductors,yankowitz2019tuning,oh2021evidence,tanaka2025superfluid,banerjee2025superfluid}, correlated insulator behavior \cite{cao2018correlated,codecido2019correlated,inbar2023quantum,nuckolls2023quantum,tian2024dominant,krishna2025terahertz}, and topological states:
quantum anomalous Hall (QAH) insulators \cite{sharpe2019emergent,serlin2020intrinsic}, Chern insulators \cite{nuckolls2020strongly,wu2021chern,choi2021correlation,stepanov2021competing}, and fractional Chern insulators \cite{xie2021fractional}.
These observations have motivated considerable theoretical efforts to
model the interplay between the large-scale atomic reconstruction and the exotic electronic properties arising from isolated flat bands.
The majority of computational studies make use of continuum models ($\kp$) \cite{dos2007graphene,Bistritzer2011,mele2011band,lopes2012continuum,carr2019exact,fang2019angle,guinea2019continuum,kang2023pseudomagnetic,vafek2023continuum,miao2023truncated,ezzi2024analytical}, tight-binding models \cite{trambly2010localization,jung2014accurate,fang2016electronic,carr2017twistronics}, and more minimal phenomenological models \cite{zou2018band,po2018origin,koshino2018maximally,kang2018symmetry,po2019faithful,tarnopolsky2019origin,carr2019exact,carr2019derivation,bernevig2021twisted,song2022magic,bennett2024twisted} due to the prohibitive cost of simulating twisted bilayers at small angles with first-principles methods such as density functional theory (DFT).
While these approaches yield valuable insights, they often lack atomic-scale structural and electronic features, which can be important for comparison with high resolution atomically resolved experiments \cite{inbar2023quantum,nuckolls2023quantum}.

A small number of first-principles DFT studies on tBLG have been performed \cite{uchida2014atomic,lucignano2019crucial,yananose2021chirality,ezzi2025universal}, although the scope is typically limited:
either the twist angle is restricted to relatively large values, or the properties considered are those that can already be described by simple theoretical models, such as structural relaxation and electronic band structure.
Earlier studies \cite{uchida2014atomic} were performed before experimental breakthroughs on magic-angle tBLG, at a time when many of its properties were not widely understood.
Additionally, many studies use plane-wave DFT codes \cite{lucignano2019crucial,yananose2021chirality}, which are not as efficient as codes which use a local basis, making the simulation of smaller twist angles difficult.

The recent development of optimized local basis sets for graphene \cite{bennett2025accurate} have made it possible to calculate three-dimensional atomistic wavefunctions for tBLG with small twist angles \cite{zhu2025wavefunction}.
This made it possible to calculate the textures of the low-energy wavefunctions of tBLG in real space, recently experimentally measured \cite{nuckolls2023quantum}, which cannot be done with continuum models that necessarily average out the atomic structure.
In our previous work we studied the evolution of the wavefunction character with twist angle and interaction strength, and identified a new phase transition where the flat bands exchange wavefunction character \cite{zhu2025wavefunction}.
Here, we present a more comprehensive first-principles study of tBLG for a wide range of angles, showcasing the capabilities of local orbital DFT for calculations involving large moir\'e supercells.
In particular, we compare the atomic structure and electronic band structure obtained from DFT with a local basis set using the {\sc siesta} code \cite{soler2002siesta,garcia2020siesta} to plane-wave DFT using the {\sc vasp} code \cite{Kresse1996,Kresse1996a}, finding excellent agreement for an angle of 2.45$^\circ$.
We also compare the lattice relaxation obtained from geometry relaxations using DFT to an elastic continuum model \cite{carr2018relaxation}, finding perfect agreement down to 0.987°.
We compare the electronic properties obtained from DFT to an exact $\kp$ model \cite{carr2019exact}.
While the two methods give qualitatively similar results, we find that there is a small twist angle offset between the DFT and exact $\kp$ calculations.
In the $\kp$ calculations, quantities such as the Fermi velocity and band width reach a minimum at the magic angle and increase again below.
The DFT bands do not exhibit these minima for the angles considered, suggesting they may occur at smaller angles.
Finally, we provide more details of the wavefunction character first presented in Ref.~\cite{zhu2025wavefunction}, and show that the Dirac cones in the K and K' valleys have the same chirality, a characteristic feature of tBLG \cite{zou2018band}.

\begin{figure*}[t]
\centering
\includegraphics[width=1.0\textwidth]{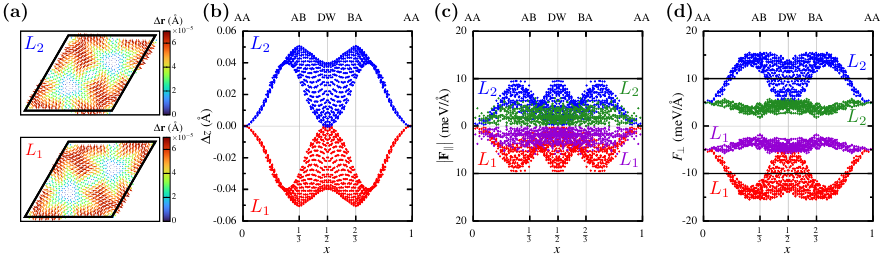}
\caption{%
Comparison of the atomic structure and forces from DFT calculations using {\sc siesta} and {\sc vasp}, for tBLG with a twist angle of $\th = 2.45^{\circ}$ and a monolayer lattice constant of $a_0 = 2.46$ Å.
{\bf (a)-(b)} The difference in positions obtained from {\sc siesta} and {\sc vasp}. 
{\bf (a)} the in-plane direction ($\D\rv$). 
The in-plane displacements due to relaxation are on the order of $10^{-2}$ \AA~for both codes, and the difference $\D\rv$ is less than $1\%$ of the relaxation.
{\bf (b)} the out-of-plane direction ($\D z$), projected onto the unit cell diagonal, where the top layer is shown in blue and the bottom layer in red.
{\bf (c)-(d)} The forces obtained from {\sc vasp} using the relaxed coordinates determined by {\sc siesta} (red/blue) and {\sc vasp} (green/purple) are shown for: {\bf (c)} the in-plane direction ($\bvec{F}_{\parallel}$); {\bf (d)} the out-of-plane direction ($F_{\perp}$), projected onto the unit cell diagonal.
The horizontal black lines show the force tolerance used in the {\sc vasp} relaxation (10 meV/\AA).
}
\label{fig:vasp-siesta}
\end{figure*}

\section{First-principles calculations}

First-principles density functional theory (DFT) calculations were performed using the {\sc siesta} \cite{soler2002siesta} code, version 5.0, with norm-conserving \cite{norm_conserving} {\sc psml} pseudopotentials \cite{psml}, obtained from Pseudo-Dojo \cite{pseudodojo}.
{\sc siesta} uses a basis of numerical atomic orbitals (NAOs) \cite{junquera2001numerical}, and basis sets optimized for graphene from Ref.~\cite{bennett2025accurate} were used in all calculations.
In order to keep the calculations efficient for large twisted supercells, we use a basis of single-$\zeta$ polarized (SZP) orbitals including $2s$ and $2p$ valence orbitals, and $3d$ polarization orbitals for each carbon atom. With one radial basis function each this results in 9 basis functions per C atom.
For calculations in `configuration space' \cite{carr2018relaxation} used to parametrize the lattice relaxation model in the next section, the more accurate double-$\zeta$ polarized $+f$ (DZPF) basis set was also used for comparison.
The DZPF basis set includes two radial basis functions per valence orbital, the $3d$ polarization orbital, as well as a higher $4f$ ($l+2$) polarization orbital for additional angular flexibility, resulting in 20 basis functions for each C atom.

\begin{figure*}[t]
\centering
\includegraphics[width=\linewidth]{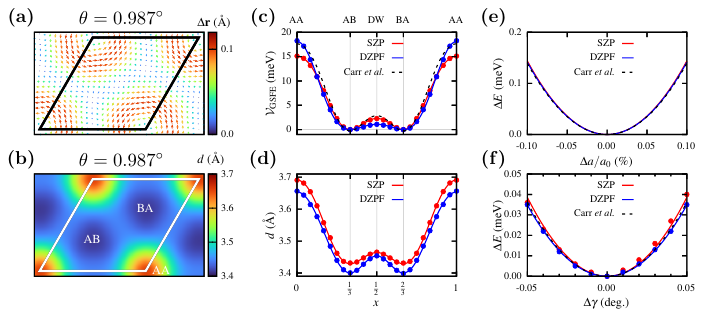}
\caption{%
{\bf (a)-(b)} Lattice relaxation in tBLG from DFT, for $\th = 0.987^{\circ}$:
{\bf (a)} the in-plane displacement $\D\rv$ of the top layer, obtained as the difference in positions between the rigid and relaxed structures, and
{\bf (b)} layer separation $d$.
{\bf (c)-(d)} GSFE and interlayer distance calculated using several different basis sets in {\sc siesta}:
{\bf (c)} the stacking energy, and
{\bf (d)} the interlayer separation were calculated for untwisted graphene bilayers by sliding one layer over the other along the unit cell diagonal, using the SZP basis (red) and the DZPF basis (blue).
The dashed black line shows the GSFE obtained using the fitting parameters in Ref.~\cite{carr2018relaxation} (Carr \textit{et al.}).
The circles represent results from DFT calculations, and the lines show fits to the data.
{\bf (e)-(f)} Elastic properties of monolayer graphene obtained from derivatives of total-energy curves: {\bf (e)} energy versus lattice constants changes used to calculate the bulk modulus; 
{\bf (f)} energy versus shear angle used to calculate the shear modulus.
For the bulk modulus, the lattice constant $a$ was perturbed about its equilibrium value $a_0$.
For the shear modulus, the angle $\g$ between the in-plane lattice vectors was modified about its equilibrium value $\g_0 = 60^{\circ}$.
}
\label{fig:lattice-relax}
\end{figure*}

The PBE exchange-correlation functional \cite{pbe} was used in all calculations, and a DFT-D3 dispersion correction \cite{grimme2010consistent} was included in order to treat the long range interactions between the layers.
A single $\kv$-point $(\G)$ was used in all self-consistent calculations and the real space grid was determined using an energy cutoff of 300 Ry.
Calculations were performed using the MRRR solver for geometry relaxations, including only the first 20 conduction bands above the Fermi level.
The SCF was converged until the relative changes in the density matrix were below $10^{-4}$.

Commensurate twisted bilayer structures were generated for angles down to $\th = 0.987^{\circ}$.
Starting with two flat layers with lattice constant $a_0 = 2.46$ Å and a layer separation of $d=3.35$~\AA, the atomic positions were relaxed until the forces on all atoms were below 10 meV/\AA.
The center of mass was fixed to the middle of the cell to prevent the bilayer from drifting from its initial position.

After the relaxed geometries were obtained, calculations were performed to obtain the electronic band structures, using the default solver.
The bands were obtained along the path \mbox{K--$\G$--M--K}, with 10 points along each segment.

For comparison, similar relaxation and band structure calculations were carried out for a twist angle of 2.45$^\circ$ using {\sc vasp} with an energy cutoff of 400 eV.
Van der Waals (vdW) corrections using geometry-dependent dispersion coefficients were included based on the zero-damping DFT-D3 method \cite{grimme2010consistent}. 
Using a single $\kv$-point ($\G$), the ionic positions were relaxed until the magnitude of all forces was less than 10 meV/Å. 
The ground-state energies and band structures were computed from the relaxed positions using a $\G$-centered $3\times 3$ $\kv$-point grid which has five $\kv$-points in the irreducible Brillouin zone, including the high-symmetry point K.

Fig.~\ref{fig:vasp-siesta} shows a comparison of the atomic structure and forces obtained from {\sc siesta} and {\sc vasp} calculations for $\th = 2.45^{\circ}$.
The differences in the in-plane and out-of-plane relaxed coordinates are shown in Figs.~\ref{fig:vasp-siesta} (a) and (b), respectively.
The differences are very small, less than $10^{-4}$ Å for the in-plane coordinates.
Interestingly the differences in relaxed coordinates have some subtle structure.
The reason for this is likely that the two codes treat the interlayer interactions slightly differently, leading to slightly different estimations of the in-plane lattice relaxation and out-of-plane corrugation.
Figs.~\ref{fig:vasp-siesta} (c) and (d) respectively show the in-plane ($F_\parallel$) and out-of-plane ($F_\perp$) forces at atomic positions projected onto the unit cell diagonal.
{\sc vasp} was used to calculate the forces for two different relaxed bilayers, both with the same commensurate twist angle of $\theta=2.45^\circ$.
The first structure was relaxed using {\sc vasp} until the magnitude of the forces on all atoms were below 10 meV/$\AA$.
The parallel and perpendicular components of the forces for that structure, plotted in green and purple for the two layers, naturally fall between the black lines that indicate the force tolerance.
The second structure was relaxed to the same force tolerance using {\sc siesta} before the residual forces were calculated in a final pass using {\sc vasp}.
Not surprisingly, the residual forces for this second structure, plotted in red and blue, are slightly larger from the perspective of the {\sc vasp} code.
In particular, the biggest difference is in the equilibrium layer separation, as evidenced by the larger forces trying to separate the layers of the {\sc siesta}-relaxed structure.
The differences in the out-of-plane structure reflect the challenge of handling the weak interlayer van der Waals forces with DFT. These results illustrate that the atomic relaxation of tBLG obtained from \textsc{siesta} using an optimized local orbital basis set are very similar to those obtained using highly converged plane wave calculations.

\section{Results
\label{sec:results}
}


\subsection{Lattice relaxation}

\begin{figure}[t]
\centering
\includegraphics[width=\columnwidth]{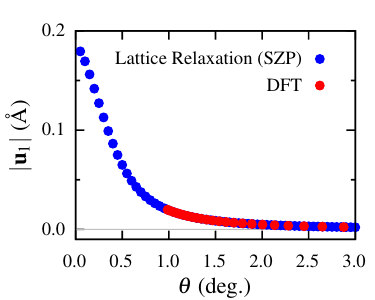}
\caption{%
The absolute value $\left | \u_1 \right|$ of the first Fourier component of the displacement field (Eq.~\eqref{eq:u-displacement}) as a function of twist angle.
The values obtained from lattice relaxation calculations are shown in blue, which were obtained by numerically minimizing Eq.~\eqref{eq:V-tot} using the SZP parameters in Table \ref{tab:GSFE}.
The red circles show the corresponding Fourier components of the relaxed DFT structure, obtained by fitting the in-plane displacement to Eq.~\eqref{eq:u-displacement} for each twist angle.
}
\label{fig:u1}
\end{figure}

\begin{figure*}[t]
\centering
\includegraphics[width=0.9\textwidth]{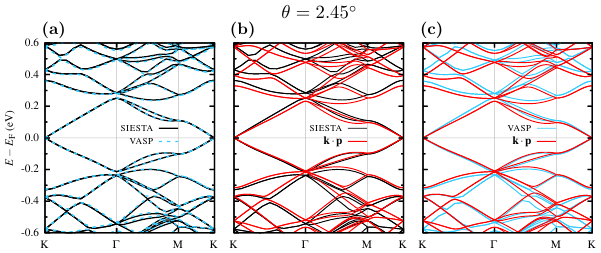}
\caption{%
Comparison of electronic band structure of tBLG for $\theta = 2.45^\circ$ calculated using \textsc{siesta}, \textsc{vasp}, and the $\kp$ model.
The black lines show the bands from DFT calculations with {\sc siesta}, the blue lines show the bands from DFT calculations with {\sc vasp}, and the red lines show the bands obtained using the $\kp$ model.
}
\label{fig:bands-kp-siesta-vasp}
\end{figure*}

\begin{figure*}[t]
\centering
\includegraphics[width=\textwidth]{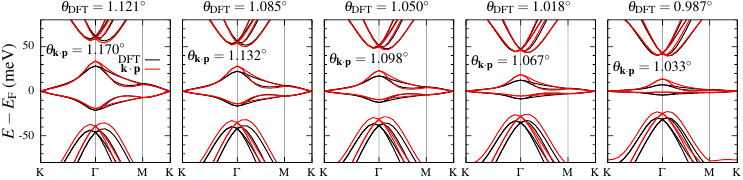}
\caption{%
Electronic band structure of tBLG for commensurate twist angles near the magic angle.
The black lines show the bands from DFT calculations with {\sc siesta} and the red lines show the bands obtained using the $\kp$ model. 
A small shift of the twist angle used in the $\kp$ calculation, $\th_{\kp} = \th_{\rm DFT} + \D\th$ where $\D\th \approx 0.05^{\circ}$, improves the agreement with the DFT bands.
}
\label{fig:bands-kp-DFT}
\end{figure*}

\begin{figure*}[t]
\centering
\includegraphics[width=\textwidth]{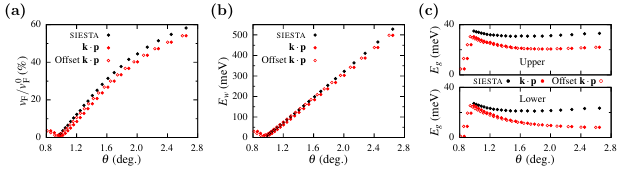}
\caption{%
{\bf (a)} Fermi velocity as a function of twist angle from {\sc siesta} (black) and the exact $\kp$ model (red), as a percentage of the Fermi velocity of graphene, $v^0_{\rm F} = 1\times 10^{6}$ m/s. 
For the $\kp$ model, filled circles correspond to initial twist angles $\theta_{\kp}$ while open circles correspond to $\kp$ twist angles that have been shifted by $\Delta \theta \approx 0.05^\circ$ towards smaller angles.
{\bf (b)} Band width $E_{w}$ (meV) at $\G$ as a function of twist angle from {\sc siesta} (black) and the exact $\kp$ model (red).
{\bf (c)} Band gaps $E_{g}$ (meV) between the flat and dispersive bands at $\G$ as a function of twist angle from {\sc siesta} (black) and the exact $\kp$ model (red).
The upper (lower) panel corresponds to the gap above (below) the Fermi level.
}
\label{fig:fermi-velocity}
\end{figure*}

\begin{figure}[t]
\centering
\includegraphics[width=1.0\columnwidth]{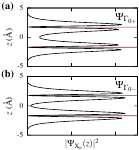}
\caption{%
Probability density of the {\bf (a)} $\Gfp$ and {\bf (b)} $\Gfm$ wavefunctions of magic-angle tBLG (1.08$^\circ$) in the out-of-plane direction, averaged over the in-plane directions.
The red lines indicate the average vertical positions of the two layers.
}
\label{fig:WF-z}
\end{figure}

\begin{figure*}[p]
\centering
\includegraphics[width=1.00\textwidth]{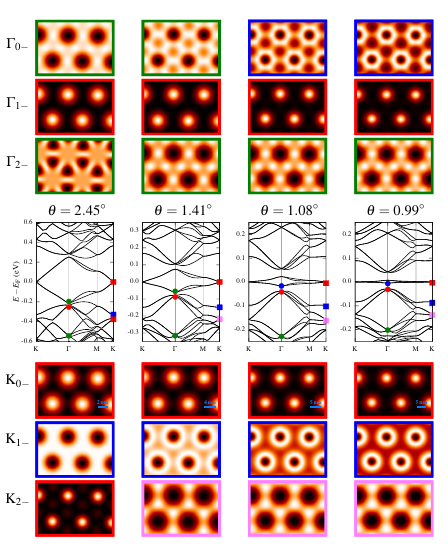}
\caption{%
Electronic band structures and wavefunctions as a function of twist angle $\th$.
The points $\G_{n -}$ (K$_{n -}$), where $n = 0,1,2$, are shown are highlighted on the band structures by the circles (squares), and the corresponding wavefunctions are shown above (below) the band structures.
The color of the markers on the band structures and the borders of the wavefunctions indicates the character of the wavefunctions as described in the text.
}
\label{fig:WF}
\end{figure*}

\begin{figure*}[t]
\centering
\includegraphics[width=\textwidth]{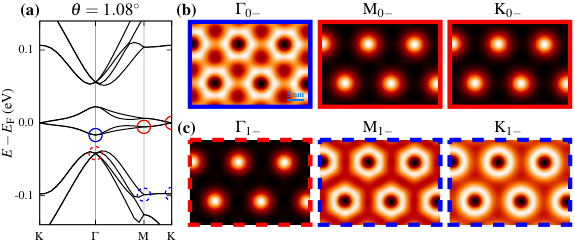}
\caption{%
{\bf (a)} Band structure of tBLG at the magic angle ($\theta = 1.08^{\circ}$). 
{\bf (b,c)} Moiré-scale wavefunctions in {\bf (b)} the lower flat band, and {\bf (c)} the dispersive valence band, at the high-symmetry $\kv$-points $\G$, M and K.
The panel borders match the respective location of the wavefunctions in the band structure, with solid (dashed) lines representing flat (dispersive) bands, and red and blue indicating wavefunctions that have AA and AA$_z$ character, respectively.
}
\label{fig:inversion}
\end{figure*}

Here we compare the relaxation of tBLG from first-principles to the relaxation from a continuum elastic model; the latter type of model is widely used to estimate the structure in moiré materials \cite{carr2018relaxation}.
The in-plane and out-of-plane lattice relaxation in tBLG with $\th = 0.987^{\circ}$ from DFT are shown in Figs.~\ref{fig:lattice-relax} (a) and (b), respectively. 
Fig.~\ref{fig:lattice-relax} (a) shows the in-plane displacement $\D\rv$ in the top layer as a result of atomic relaxation, starting from two layers with a rigid twist.
Most of the displacement is concentrated in rings around the AA sites.
Fig.~\ref{fig:lattice-relax} (b) shows the layer separation $d$, defined as the difference in the out-of-plane coordinates between the top and bottom layers after geometry relaxation.
Most of the corrugation occurs around the AA sites.
For comparison to the DFT results, we performed relaxation calculations using a continuum model employing the concept of the generalized stacking fault energy (GSFE)~\cite{kaxiras1993free}.
For completeness, a derivation of the lattice relaxation continuum model is given in Appendix \ref{appendix}, following the methodology in Refs.~\cite{carr2018relaxation,bennett2022electrically,bennett2022theory,ramos2025flat}.

All of the material parameters in the total energy, namely the GSFE parameters ($\V_i$) and the elastic coefficients $B$ (bulk modulus) and $\mu$ (shear modulus), were obtained from our first-principles calculations.
The GSFE is shown in Fig.~\ref{fig:lattice-relax} (c), where the circles are results from DFT calculations, and the lines are the fits to $\VGSFE$, Eq.~(\ref{eq:V-GSFE}).
The SZP (DZPF) calculations are shown in red (blue), and the dashed black line shows the GSFE form from Carr {\it et al.}~\cite{carr2018relaxation}, all of which are in good agreement apart from minor details.
The geometry of the commensurate bilayer was also optimized for each stacking to obtain the interlayer separation as a function of stacking (corrugation).
The results are summarized in Fig.~\ref{fig:lattice-relax} (d), which includes the results from DFT as well as a fit to a smooth function similar to Eq.~\eqref{eq:V-GSFE}, with coefficients $d_i$ (Table \ref{tab:GSFE}).
Figs.~\ref{fig:lattice-relax}(e) and (f) show the change in total energy as a function of lattice constant and cell angle, respectively, as well as the fit to Eqs.~\eqref{eq:bulk-mod} and \eqref{eq:shear-mod} used to extract the bulk and shear moduli.
All these material parameters are collected in Table~\ref{tab:GSFE} and compared with the values in Ref.~\cite{carr2018relaxation}.

One way to provide a quantitative comparison of the relaxation patterns obtained from DFT and continuum models is to compare the Fourier coefficients of the displacement field $\u$.
To this end, we have calculated this displacement field (and its Fourier expansion) for a wide range of twist angles with a fine grid,  
using the parameters in Table \ref{tab:GSFE} for the continuum relaxation.
The magnitude of the leading Fourier component, $\left| \u_1\right|$, is shown in Fig.~\ref{fig:u1} in blue dots, from $\th = 3^{\circ}$ down to $\th = 0.05^{\circ}$ in steps of $0.05^{\circ}$.
Starting at around the magic angle ($\sim 1.08^{\circ}$), 
$\left| \u_1\right|$ increases rapidly with decreasing twist angle, indicating significant lattice relaxation.
The displacement fields after DFT relaxation were fit to Eq.~\eqref{eq:u-displacement} for commensurate angles down to $\th=0.987^{\circ}$, and the values of $\left| \u_1\right|$ are shown in Fig.~\ref{fig:u1} as red dots.
In this comparison, 
the agreement between DFT relaxation and continuum relaxation is perfect for angles as small as $\th=0.987^{\circ}$, the limit we could reach with DFT calculations. 

\subsection{Electronic Properties}

Here we compare the electronic properties of tBLG obtained from our first-principles \textsc{siesta} calculations to other methods.

Fig.~\ref{fig:bands-kp-siesta-vasp} (a) shows a comparison of the band structures at $\theta = 2.45^\circ$ obtained from \textsc{siesta} and \textsc{vasp}, with near perfect agreement for the flat bands and nearest dispersive bands.
This result illustrates that in addition to the structural properties, the electronic properties obtained from {\sc siesta} calculations using an optimized basis set give comparable results to highly converged plane-wave calculations like {\sc vasp}, further supporting the argument that it is possible to perform accurate calculations for larger systems with smaller twist angles using the computationally more efficient local orbital methodology.

We next compare the electronic structure calculated using \textsc{siesta} to the most accurate $\kp$ model \cite{carr2019exact} that exactly reproduces the results of \textit{ab initio} tight-binding models \cite{jung2014accurate,fang2016electronic}. Fig.~\ref{fig:bands-kp-siesta-vasp} (b) and (c) compare the $\kp$ band structure to the \textsc{siesta} and \textsc{vasp} band structures, respectively, for $\theta = 2.45^\circ$. The $\kp$ band structure shows slight differences in the Fermi velocity, band width, and band gaps, compared to the DFT band structures calculated from \textsc{siesta} and \textsc{vasp}. However, we note that the $\mathbf{k}\cdot \mathbf{p}$ model is less accurate for angles far above the magic angle. Finally, in Fig.~\ref{fig:bands-kp-DFT} we compare the \textsc{siesta} band structures with the corresponding $\kp$ bands for a selection of commensurate angles near the magic angle.
The two calculations exhibit very close agreement if the twist angle used in the $\kp$ calculations is shifted slightly,
$\th_{\kp} = \th_{\rm DFT} + \D\th$, with $\D\th \approx 0.05^{\circ}$.
This merely states that the definition of the `magic range' is slightly different in the two approaches, due to the differences in computational methods as well as different ways of representing the interlayer interactions.
We elaborate on this issue in the Discussion.  
The twist-angle shifts used in Fig.~\ref{fig:bands-kp-DFT} 
give very good agreement between the upper dispersive bands, and improved agreement between the flat bands, 
while a small energy offset persists for the lower dispersive bands.

In order to further investigate the difference between the DFT and $\kp$ bands, 
Fig.~\ref{fig:fermi-velocity} shows the Fermi velocity, band width, and band gaps calculated using {\sc siesta} for angles as low as $\th=0.987^\circ$.
The $\kp$ bands were sampled for the same commensurate angles and three additional smaller angles.
For all three features, the DFT and $\kp$ calculations exhibit the same twist angle dependence, but with the small offset in twist angle mentioned above.

The Fermi velocity, $v_{\rm F} = \left.\frac{1}{\hbar}\frac{dE}{d k}\right|_{\rm K}$, is the slope of the bands about the Dirac node, which we calculate along the path K--$\G$.
The Fermi velocity, normalized by the Fermi velocity of monolayer graphene, $v_{\rm F}^0 = 10^6$ m/s, is shown in Fig.~\ref{fig:fermi-velocity} as a function of twist angle.
The $\kp$ Fermi velocity reaches a minimum and then starts increasing again as the twist angle decreases below the magic angle \cite{carr2019exact,bennett2024twisted}.
The Fermi velocity extracted from the {\sc siesta} calculations does not show a minimum within the range of twist angles we simulated. 

In Fig.~\ref{fig:fermi-velocity} (b) we show the band width, defined as the energy difference between the upper and lower flat bands at $\G$, and in Fig.~\ref{fig:fermi-velocity} (c) we show both upper and lower band gaps, the energy differences between the flat and dispersive bands at $\G$.
There is generally good agreement between the band width from {\sc siesta} and the $\kp$ model,
although again we see that the {\sc siesta} band width does not reach a minimum.
Considering the band gaps, the upper gap is larger than the lower gap for all twist angles in both calculations, reflecting the particle hole asymmetry of the bands.
For the $\kp$ bands, the gaps abruptly go to zero below the magic angle, as the dispersive bands move closer to the Fermi level and eventually intersect with the flat bands \cite{bennett2024twisted}.

Lastly, our calculated band widths are in good agreement with experimental estimates \cite{li2024evolution,lisi2021observation}, shown in Table \ref{tab:exp-bandwidth}.

\begin{table}[t]
\begin{tabular}{|c|c|c|}
\hline
\hline
\multicolumn{1}{|c|}{\multirow{2}{*}{$\theta$ $(^\circ)$}} & \multicolumn{2}{c|}{Band width (meV)} \\
\cline{2-3} 
& \textsc{siesta} (lower flat band) & Experiment \\
\hline
$2.60$ & $238$ & $215 \pm 30$ \cite{li2024evolution} \\
$1.93$ & $136$ & $111 \pm 24$ \cite{li2024evolution} \\
$1.34$ & $51$ & $30 \pm 15$ \cite{lisi2021observation} \\
$1.31$ & $46$ & $45 \pm 15$ \cite{li2024evolution} \\
$1.07$ & $15$ & $11 \pm 10$ \cite{li2024evolution} \\
\hline
\hline
\end{tabular}
\caption{Band width of the lower flat band calculated using \textsc{siesta} and measured by nano-ARPES experiments. The \textsc{siesta} values are interpolated from the nearest commensurate angles.
}
\label{tab:exp-bandwidth}
\end{table}

\subsection{Moiré-scale wavefunctions}

An advantage of explicit DFT calculations with all the atomic degrees of freedom is that they provide realistic single-particle wavefunctions that can elucidate the physics and allow detailed comparison with experiments, such as scanning tunneling microscope images ~\cite{nuckolls2023quantum}.  
Accordingly, we turn next to an analysis of the character of the DFT wavefunctions for tBLG, initially reported in Ref.~\cite{zhu2025wavefunction}.

In {\sc siesta}, atomic-scale wavefunctions are represented as linear combinations of the local (non-orthogonal) basis orbitals $\phi_{\mu}$:
\beq{}
\Psi_{n\kv}(\rv) = \sum_{\mu}c_{n\kv,\mu}\phi_{\mu}(\rv)
\eep
We obtained real-space wavefunctions $\Psi_{n\kv}(\rv)$ on a three-dimensional grid for the low-energy bands $n$ at the high-symmetry $\kv$ points $\Gamma$, M, and K, in the moiré BZ (mBZ).
In the following we use the notation introduced in Ref.~\cite{zhu2025wavefunction} to refer to the wavefunctions: X$_{n\pm}$, where $\mathrm{X}=\G,\ \mathrm{K},\ \mathrm{M}$ are the high-symmetry $\kv$-points in the mBZ, and $n = 0,1,2,\ \ldots$ is the band index: $n=0$ refers to the flat bands, $n=1$ to the first dispersive bands, and so on.
The sign indicates whether the bands are above ($+$) or below ($-$) the charge neutrality point.
For example, $\Gfp$ and $\Gfm$ refer to the wavefunctions of the upper and lower flat bands at $\G$, respectively, and $\Kdp$ and $\Kddp$ refer to the wavefunctions at K of the first and second dispersive bands above charge neutrality.

The structure of the wavefunctions in the out-of-plane direction, obtained by averaging the probability density over the plane of the layers, is shown in Fig.~\ref{fig:WF-z} for two representative cases, namely $\Psi_{\Gamma_{0+}}$ and 
$\Psi_{\Gamma_{0-}}$, each of which corresponds to a 
two-fold degenerate state at the center of the mBZ 
belonging to the flat band above and below the charge neutrality point, respectively.
The probability densities have peaks on either side of each graphene layer, which is expected because each one is a linear combination of the $2p_z$ valence electrons of the C atoms within each graphene layer.

What matters most for the behavior of electrons in tBLG is the in-plane (two-dimensional) character of the wavefunctions.
In order to obtain a two-dimensional representation of $\Psi_{n\kv}(\rv)$, we show the modulus squared of the wavefunction, summed in the out-of-plane direction $z$:
\beq{eq:psi_z_average} 
{\vert\Psi_{n\kv}(x, y)\vert^2 = \sum_{z} \vert\Psi_{n\kv}(x, y, z)\vert^2}
\eep
For degenerate states, we add the moduli squared of the degenerate wavefunctions.
The averaging over values of $z$ introduced in Eq.~\eqref{eq:psi_z_average} obscures the behavior of the wavefunctions in each layer and each sublattice ($A$ or $B$) of the honeycomb lattice. This behavior is important in characterizing certain symmetries of the wavefunctions and is discussed in Section~\ref{sec:symmetry}. 

Wavefunctions were obtained at the high symmetry $\kv$-points $\G$, K and M, for the flat bands and neighboring dispersive bands.
The bare wavefunctions, which we refer to as the microscopic wavefunctions, have intricate details on the scale of the monolayer graphene lattice constant \cite{zhu2025wavefunction}.
To visualize the moiré-scale features of the atomic-scale wavefunctions, we perform a convolution of $\vert\Psi_{n\kv}(x, y)\vert^2$ with a symmetric two-dimensional Gaussian filter, obtaining a (modulus squared) ``moiré-scale" probability density $\vert\widetilde{\Psi}_{n\kv}\vert^2$ given by
\beq{}
\vert\widetilde{\Psi}_{n\kv}\vert^2(x, y) \equiv (\vert\Psi_{n\kv}\vert^2 * \N)(x, y) = \sum_{x', y'} \vert \Psi_{n\kv}(x', y')\vert^2 \N_{x,y}(x',y')
\eep
Here, $\N_{x,y}(x',y')$ denotes the probability density function of a Gaussian with mean $(x, y)$ and covariance matrix $\sigma^2 I_{2 \times 2}$ evaluated at the point $(x', y')$. 
In practice, we sum over all $(x', y')$ such that $\Vert (x', y') - (x, y) \Vert \leq 3\sigma$. 
We choose $\sigma$ large enough such that we can clearly extract the moiré-scale features from the atomic-scale wavefunctions. 
Generally, we choose larger values of $\sigma$ for smaller $\theta$ because the moiré scale is larger (see Ref.~\cite{zhu2025wavefunction} for additional details).

The moiré-scale wavefunctions can broadly be separated into four categories, reflecting different features on the moiré scale, based on the notation introduced in \cite{carr2019exact}:
``AA'' wavefunctions, localized at the AA sites, forming a triangular lattice (red border in Figs.~\ref{fig:WF} and \ref{fig:inversion}), 
``AA$_z$'' wavefunctions, which form rings around the AA sites (blue border);
``DW'' wavefunctions, with weight along the domain walls, forming a Kagome lattice (green border), and
``AB/BA'' wavefunctions with weight in the AB/BA domain centers, forming a honeycomb lattice (pink border).

Fig.~\ref{fig:WF} shows the electronic band structure for a selection of twist angles between $2.45^{\circ}$ and $0.99^{\circ}$,
with colored markers indicating the category of the moiré-scale wavefunctions of selected bands at high-symmetry points in the BZ.
The corresponding in-plane probability density is shown above and below the band structures for states at $\G$ and K, respectively, with the colored border matching the wavefunction category.
We observe that the wavefunctions evolve continuously with twist angle.
For large angles, $\th = 2.45^{\circ}$, the wavefunction at $\Gfm$ has DW character. 
As the twist angle decreases, the density becomes less localized around the AB/BA domains and DWs and more around the AA sites, resulting in a continuous change to an $\mathrm{AA}_z$ wavefunction.
In the first dispersive bands ($\Gdm$), the wavefunctions exhibit AA character for all twist angles, becoming more localized as the twist angle decreases.
In the second dispersive bands ($\Gddm$), the wavefunctions exhibit DW character, which emerges at around $\th = 2.45^{\circ}$ and persists down to $\th = 0.99^{\circ}$. 
The K wavefunctions always have AA character in the flat bands, which becomes more localized as the twist angle decreases.
In the first dispersive bands, the K wavefunctions have AA$_z$ character, and the second dispersive bands have AB/BA character at smaller angles.

\subsection{Symmetry properties}
\label{sec:symmetry}

\begin{figure*}[t]
\centering
\includegraphics[width=0.9\linewidth]{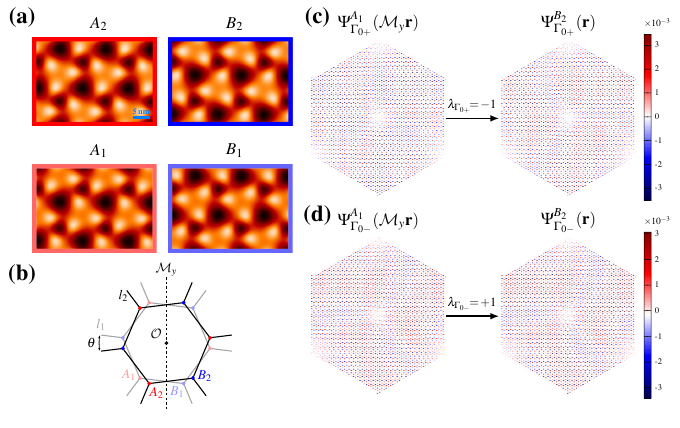}
\caption{%
{\bf (a)} Probability density of the sublattice-projected wavefunctions of magic-angle tBLG at $\Gfm$.
{\bf (b)} Sketch indicating layers $l_1$ (black) and $l_2$ (gray), sublattices $A$ (red) and $B$ (blue), and $\mathcal{M}_y$ symmetry axis.
{\bf (c)} $\M_y$ transformation of the projected microscopic wavefunctions of magic-angle tBLG for the upper flat band ($\Gfp$).
The $\M_y$ mirror operation is applied to $\Psi^{A_1}_{\Gfp}$.
The inverted colors show that the upper flat band has a mirror eigenvalue of $\la_{\Gfp}\! = -1$.
{\bf (d)} The same as {\bf (c)}, but for the lower flat band $\lb\Gfm\rb$.
In this case, $\la_{\Gfm}\! = +1$.
}
\label{fig:symmetry}
\end{figure*}

We next examine the symmetry properties of the microscopic wavefunctions in tBLG.
As we can see from Fig.~\ref{fig:inversion}, 
the flat and dispersive bands swap moir\'e-scale character near $\G$.
The band inversion and fragile topology in tBLG \cite{zou2018band,po2019faithful,ahn2019failure,song2019all}, which prevent an atomic Wannier representation using four or fewer bands \cite{po2019faithful}, are intimately related to the chirality of the wavefunctions.
A signature of the fragile topology is that the wavefunctions have the {\it same} chirality around the Dirac nodes at the K and K$'$ points, in contrast to single-layer graphene where the two Dirac nodes exhibit opposite chirality \cite{zou2018band}.
We can check the relative signs of the chiralities by checking the symmetry properties of the microscopic wavefunctions.
This approach was used in Ref.~\cite{zhu2025wavefunction} to characterize a phase transition as the twist angle is reduced below the magic angle.
For completeness, we briefly summarize the method below.

First, we calculate projections of the wavefunctions $\Psi_{n\kv}$ onto the basis functions of the individual layers and sublattices $X_l$, where $X=A,B$ for the sublattice and $l=1,2$ for the layer index:
\beq{}
\Psi_{n\kv}^s(\rv) = \sum_{\mu \in s} c_{n\kv, \mu} \phi_{\mu}(\rv)
\eec
where $s = A_1, A_2, B_1, B_2$, representing the carbon atoms at the $A$ and $B$ sites of layers 1 or 2.
We visualize these wavefunctions on the moiré scale using the same convolution procedure for the full wavefunctions described in the previous section.
Examples of sublattice-projected wavefunctions are shown in Fig.~\ref{fig:symmetry} (a).

We then check how these wavefunctions transform under the symmetry $\M_y$, which is equivalent to a $180^{\circ}$ rotation about an axis parallel to the layers, i.e., $\M_y\rv = \M_y(x,y,z) = (-x,y,-z)$ (see Fig.~\ref{fig:symmetry} (b)).
We note that although $C_{2y}$ would be a more accurate label for this symmetry, we use $\mathcal{M}_y$ to be consistent with previous literature \cite{zou2018band}.
We analyze the transformation of the $\Gfpm$ \textit{microscopic} wavefunctions under the valley-preserving $\M_y$ symmetry, which has the effect of exchanging the layers and sublattices and acts on the sublattice-projected wavefunction as follows \cite{zou2018band,fang2019angle}:
\beq{eq:mirror-eigenvalue}
\M_y
\begin{pmatrix} 
\Psi^{A_1}(\rv) \\ 
\Psi^{B_1}(\rv) \\ 
\Psi^{A_2}(\rv) \\ 
\Psi^{B_2}(\rv) 
\end{pmatrix}
=
\begin{pmatrix} 
\Psi^{B_2}(\M_y \rv) \\ 
\Psi^{A_2}(\M_y \rv) \\ 
\Psi^{B_1}(\M_y \rv) \\ 
\Psi^{A_1}(\M_y \rv) 
\end{pmatrix} 
=
\la
\begin{pmatrix} 
\Psi^{A_1}(\rv) \\ 
\Psi^{B_1}(\rv) \\ 
\Psi^{A_2}(\rv) \\ 
\Psi^{B_2}(\rv)
\end{pmatrix}
\eec
Here $\la = \pm 1$ is the $\mathcal{M}_y$ eigenvalue. 
If the wavefunctions at $\Gfp$ and $\Gfm$ have opposite eigenvalues, the Dirac nodes at K and K$'$ have the same chirality, and vice versa \cite{zou2018band}.

Fig.~\ref{fig:symmetry} (c) shows sublattice projections of the {\it microscopic} wavefunctions at $\Gfpm$. 
The projection onto $A_1$ after the $\M_y$ mirror operation has been applied is compared to the corresponding component in Eq.~\eqref{eq:mirror-eigenvalue}, namely $B_2$.
We see that $\la_{0+} = -1$ and $\la_{0-} = +1$: the wavefunctions have opposite $\mathcal{M}_y$ eigenvalues, and thus the Dirac nodes at K and K$'$ have the same chirality.
This confirms that our first principles calculations yield the expected band topology of tBLG.

\section{Discussion and Conclusions}

In this work we have presented a comprehensive first-principles study of tBLG spanning a wide range of twist angles, including those at and below the magic angle.
Using DFT with a local basis, with optimized basis sets, we have obtained fully relaxed atomic geometries and resulting electronic structures for supercells containing tens of thousands of atoms, at a level of detail inaccessible with other theoretical approaches.

Through direct comparison with plane-wave DFT calculations, we have validated the accuracy of our local basis calculations.
Our atomic relaxations are in perfect agreement with continuum elastic models down to the smallest angles considered.
This suggests that a continuum model may be used to generate relaxed structures which can be used to more efficiently simulate even smaller angles.
While the electronic properties from our DFT calculations show similar trends to exact $\kp$ models, we find a small offset in the twist angle dependence.
The twist-angle offset is linked to differences in the Fermi velocity and the band width of the flat bands, as a function of twist angle, presented in Fig.~\ref{fig:fermi-velocity}.

These differences between the present work and previous models are most likely a result of the minimal basis set used here, the exchange-correlation (XC) functional employed in our calculations, and the treatment of the vdW interactions. 
To assess the relative importance of each factor, we compute the layer separation of bilayer graphene in various high-symmetry stacking configurations (AA, AB, and DW-like configurations) using the SZP and DZPF basis sets in \textsc{siesta} and different XC functionals and vdW corrections in \textsc{vasp}, shown in Table \ref{tab:comparisons}. 
The distance separating the two layers effectively measures the interlayer interaction strength, with stronger interlayer interactions resulting in flatter bands \cite{zhu2025wavefunction}; thus, the layer separation for different high-symmetry stackings can offer physical intuition on the differences between the \textsc{siesta} and $\kp$ bands.

The \textsc{vasp} calculations in Table \ref{tab:comparisons} show that replacing PBE with the meta-GGA SCAN functional decreases the layer separation by $\sim 0.06$ to $0.09$ \AA~for the different stacking configurations, and subsequently replacing the DFT-D3 dispersion correction with the nonlocal rVV10 vdW correction further decreases the layer separation by $\sim 0.03$ to $0.07$ \AA. 
In contrast, the \textsc{siesta} calculations in Table \ref{tab:comparisons} show that replacing the SZP basis with the DZPF basis only decreases the separation distance by $\sim 0.01$ to $0.03$ \AA. 
While it is difficult to completely isolate effects of these factors due to inherent differences in the bases and pseudopotentials between \textsc{siesta} and \textsc{vasp}, these comparisons allow us to plausibly infer that the choice of XC functional and vdW correction are both important factors influencing the interlayer interaction strength, contributing to the differences between the \textsc{siesta} and $\kp$ bands.
In particular, the $\kp$ model is parameterized based on calculations using SCAN and the rVV10 vdW correction in \textsc{vasp} \cite{fang2019angle}, which produce smaller layer separations in bilayer graphene than the SZP calculations in \textsc{siesta} and are thus consistent with the observation that the $\kp$ model produces flat bands at a slightly larger angle than the \textsc{siesta} calculations. 
These tests further demonstrate that the coupling strength between the layers plays a critical role in determining the angle at which the bands in tBLG are flattest. 

\begin{table}[t]
\centering
\begin{tabular}{|c|ccc|}
\hline
\hline
\textsc{siesta} & \multicolumn{3}{c|}{$d$ (\AA)} \\
\cline{2-4}
(basis, XC, vdW) & AA & AB & DW \\
\hline
SZP, PBE, DFT-D3 & $3.689$ & $3.438$ & $3.471$ \\
DZPF, PBE, DFT-D3 & $3.656$ & $3.402$ & $3.458$ \\
\hline
\hline
\textsc{vasp} & \multicolumn{3}{c|}{$d$ (\AA)} \\
\cline{2-4}
(XC, vdW) & AA & AB & DW \\
\hline
PBE, DFT-D3 & $3.722$ & $3.543$ & $3.568$ \\ 
SCAN, DFT-D3 & $3.661$ & $3.461$ & $3.472$ \\
SCAN, rVV10 & $3.630$ & $3.390$ & $3.437$ \\
\hline
\hline
\end{tabular}
\caption{Layer separation $d$ calculated for various stacking configurations of bilayer graphene using different basis sets in \textsc{siesta} and different XC functionals and vdW corrections in \textsc{vasp}.
}
\label{tab:comparisons}
\end{table}

Finally, we provide detailed descriptions of the wavefunction character, band inversion and fragile topology of the low-energy wavefunctions of tBLG, following on from our previous work \cite{zhu2025wavefunction}.
We show the evolution of the wavefunctions of the flat and dispersive bands as a function of twist angle, and for different high symmetry $\kv$-points.
We illustrate the band inversion near $\G$ between the flat and neighboring dispersive bands, and also calculate the symmetry properties of the sublattice-projected wavefunctions, illustrating that the two Dirac nodes in our calculations have the same chirality, as expected in tBLG.
Our first-principles approach establishes the foundation for detailed, accurate, and computationally viable simulations of vdW materials with tens of thousands of atoms, with atomic resolution.

\section*{Acknowledgments}
The authors acknowledge support from the Simons Foundation award No.~896626 and the US Army Research Office (ARO) MURI project under Grant No.~W911NF-21-0147.
A.Z.~is supported by the U.S.~Department of Energy, Office of Science, Office of Advanced Scientific Computing Research, Department of Energy Computational Science Graduate Fellowship under Award Number DE-SC0025528.
D.B.~acknowledges support from the NTU Startup Grant (Award Number 025661-00003).  
E.M.~acknowledges the hospitality
of the Harvard Physics Department, where part of this work
was done.
M.M.A.E.~acknowledges support from the Ministry of Education, Singapore (Research Centre of Excellence award to the Institute for Functional Intelligent Materials, I-FIM, project No. EDUNC-33-18-279-V12), the National Research Foundation, Singapore, under its AI Singapore Programme (AISG Award No: AISG3-RP-2022-028).

\appendix
\section{Derivation of lattice relaxation continuum model}
\label{appendix}

\begin{table*}[t]
\begin{tabular}{|c|r|r|r|r|r|r|r|r|r|r|}
\hline\hline
& $B$ (meV) & $\mu$  (meV) & $\V_0$ (meV) & $\V_1$ (meV) & $\V_2$ (meV) & $\V_3$ (meV) & $d_0$ (Å) & $d_1$ (Å) & $d_2$ (Å) & $d_3$ (Å) \\ \hline
SZP & $71929$ & $50667$ & $5.832$ & $3.458$ & $-0.262$ & $-0.115$ & $3.353$ & $0.059$  & $-0.003$ & $-0.001$ \\ 
DZPF & $70754$ & $45704$ & $5.367$ & $4.030$ & $0.223$ & $-0.010$ & $3.500$ & $0.055$  & $-0.005$ & $0.001$ \\ 
Ref.~\cite{carr2018relaxation} & $69518$ & $47352$ & $6.832$ & $4.064$ & $-0.374$ & $-0.095$ & - & - & - & - \\ 
\hline\hline
\end{tabular}
\caption{%
Fitting of material parameters used in the total energy in Eq.~\eqref{eq:V-tot}, obtained from first-principles calculations: 
the bulk modulus $B$ and shear modulus $\mu$, the GSFE Fourier components $\V_i$, as well as the interlayer separation Fourier components $d_i$ (not included in the total energy).
The values obtained using the SZP and DZPF basis sets are shown, as illustrated in Fig.~\ref{fig:lattice-relax}, and the values from Ref.~\cite{carr2018relaxation} (Carr \textit{et al.}) are shown for comparison.
}
\label{tab:GSFE}
\end{table*}

Here we derive the continuum model for lattice relaxation, following the methodology in Refs.~\cite{carr2018relaxation,bennett2022electrically,bennett2022theory,ramos2025flat}.
We minimize the total energy in configuration space:
\beq{eq:V-tot}
V_{\rm tot} = \int \left[\VGSFE(\x + \u(\x)) + \Velastic(\nabla\u)\right] \dd \x
\eec
where $\x$ denotes the local stacking, and $\u(\x)$ is the additional displacement which occurs as a result of lattice relaxation.
In configuration space, the integration is performed over a cell which contains all possible local stackings, which is the same size and shape as the monolayer primitive unit cell.
The GSFE of tBLG is given by
\beq{eq:V-GSFE}
\begin{split}
\VGSFE = &\V_{1} \left[\cos(2\pi x) + \cos{(2\pi y)} + \cos{(2\pi(x+y))}\right] + \\
          &\V_{2} \left[\cos(2\pi (x+2y)) + \cos{(2\pi (2x+y))} + \cos{(2\pi(x-y))}\right] + \\
          &\V_{3} \left[\cos(4\pi x) + \cos{(4\pi y)} + \cos{(4\pi(x+y))}\right] + \ldots
\end{split}
\eeq
in fractional coordinates.
The mapping to Cartesian coordinates $\x \to \bvec{s}$ is given by
\beq{}
\begin{gathered}
\bvec{s} = A\x, \quad
A = \begin{bmatrix} \av_1 & \av_2 \end{bmatrix}, 
\quad 
\av_1 = \begin{bmatrix} 1 \\ 0 \end{bmatrix},\quad
\av_2 = \begin{bmatrix} 1/2 \\ \sqrt{3}/2 \end{bmatrix}
\end{gathered}
\eeq
where $\av_1$ and $\av_2$ are the lattice vectors in units of the graphene lattice constant $a_0$.
For tBLG, the GSFE in Eq.~\eqref{eq:V-GSFE} is even and has an effective $\C_6$ symmetry in the continuum limit.

The elastic energy is given by
\beq{}
\Velastic = C_{ijkl}\eta_{ij}\eta_{kl}
\eec
where $C_{ijkl}$ is the linear elastic tensor \cite{bennett2022electrically}, and
\beq{}
\eta_{ij} = \frac{1}{2}\lb \partial_i u_j + \partial_j u_i\rb
\eeq
is the strain field.
The elastic energy in configuration space (fractional coordinates) is given by \cite{bennett2022electrically,bennett2022theory}
\beq{}
\begin{split}
\V_{\text{elastic}} &= \frac{\th^2}{2} \bigg\{ B\lb \partial_{x}u_{y} - \partial_{y}u_{x}\rb^2 \bigg. + \\
& \bigg. \mu \left[ \frac{4}{3}\lb \partial_{x}u_{y} + \partial_{y}u_{x} \rb^2 + \lb  \partial_{x}u_{x} - \partial_{y}u_{y}\rb^2 \right] \bigg\}
\end{split}
\eec
where $B$ and $\mu$ are the bulk and shear moduli of monolayer graphene.
The displacement field $\u$ is expanded in Fourier harmonics,
\beq{eq:u-displacement}
\u(\x) = \sum_{\gv} \u_{\gv} \exp{[i(A^{-1} \gv)\cdot\bvec{x}]}
\eec
and the total energy in Eq.~\eqref{eq:V-tot} is minimized with respect to the Fourier coefficients $\u_{\gv}$: $\nabla_{\u_{\gv}}V_{\rm tot} = 0$.

All of the material parameters in the total energy, namely the the GSFE ($\V_i$) and the elastic coefficients $B$ (bulk modulus) and $G$ (shear modulus), are parametrized using first-principles calculations.
Calculations were performed to parametrize Eq.~\eqref{eq:V-tot}, taking commensurate bilayer graphene and sliding one layer over the other, with the results summarized in Table \ref{tab:GSFE}.
Calculations were performed using both the SZP basis set, which was used to perform the tBLG calculations, as well as the more accurate DZPF basis set \cite{bennett2025accurate}.
The top layer was translated over the bottom layer along the unit cell diagonal, $\mathbf{s} = x(\av_1 + \av_2)$, which is sufficient to get a good fit to $\VGSFE$ as it contains all of the high symmetry stackings (AA, AB, DW, BA).
The geometry of the commensurate bilayer was also optimized for each stacking to obtain the interlayer separation as a function of stacking (corrugation).
The results from DFT calculations as well as the fitting of the GSFE and interlayer separation are shown in Figs.~\ref{fig:lattice-relax} (c) and (d) respectively, and are compared with the values in Ref.~\cite{carr2018relaxation} (Table \ref{tab:GSFE}).

The bulk modulus is obtained from a quadratic fit to the total energy as a function of the lattice constant $a$ for small variations about the equilibrium lattice constant $a_0$:
\beq{eq:bulk-mod}
E(a) = 2 B \lb\frac{a-a_0}{a_0}\rb^2
\eep
The shear modulus was similarly obtained from a quadratic fit to the total energy as a function of
the cell angle $\g$ between $\av_1$ and $\av_2$ for small variations about the equilibrium angle $\g_0=60^\circ$:
\beq{eq:shear-mod}
E(\g) = \mu \lb \g-\g_0 \rb^2
\eep
The results from DFT calculations as well as the fit to Eqs.~\eqref{eq:bulk-mod} and \eqref{eq:shear-mod} are shown in Figs.~\ref{fig:lattice-relax} (e) and (f) respectively, and are compared with the values in Ref.~\cite{carr2018relaxation} (Table \ref{tab:GSFE}).

\end{document}